\documentclass[]{aa}  

\usepackage{graphicx}
\usepackage{txfonts}

\usepackage{epstopdf}
\usepackage{amsmath}
\usepackage[usenames, dvipsnames]{xcolor}
\usepackage{textcomp,gensymb}
\usepackage[normalem]{ulem}
\usepackage{natbib}
\usepackage{amsmath}

\usepackage[]{hyperref}

\def\kms{{\rm km\,s^{-1}}}
\def\masyr{{\rm mas\,yr^{-1}}}
\def\mutot{\mu_{\rm tot}}
\def\mub{\mu_{b}}
\def\mul{\mu_{{\ell*}}}
\def\mud{\mu_{\delta}}
\def\mua{\mu_{{\alpha*}}}
\def\HP{HEALpix }
\def\HPf{HEALpix}
\def\L{\tilde{\Lambda}_\odot}
\def\B{\tilde{B}_\odot}
\def\Gaia{{\it Gaia }}

\begin{document} 





   \title{An all-sky proper motion map of the Sagittarius stream using \Gaia DR2}
\titlerunning{Proper motion of Sgr}
   \subtitle{}

   \author{T. Antoja
          \inst{1}
          \and
          P. Ramos\inst{1}
          \and
          C. Mateu\inst{2}
                \and
          A. Helmi\inst{3}
                \and
          F. Anders\inst{1}
           \and
          C. Jordi\inst{1}
                   \and
          J. A. Carballo-Bello\inst{4}
          }

   \institute{Institut de Ci\`{e}ncies del Cosmos, Universitat  de  Barcelona  (IEEC-UB), Mart\'{i} i Franqu\`{e}s  1, E-08028 Barcelona, Spain\\
              \email{tantoja@fqa.ub.edu}
         \and
             Departamento de Astronomía, Facultad de Ciencias, Universidad de la República, Iguá 4225, 14000, Montevideo, Uruguay
             \and
             Kapteyn Astronomical Institute, University of Groningen, NL-9747 AD Groningen, Netherlands
             \and
             Instituto de Astrofísica, Pontificia Universidad Católica de Chile, Av. Vicuña Mackenna 4860, 782-0436 Macul, Santiago, Chile
             }

   \date{Received xxxx; accepted xxxx}

 
  \abstract
   {}
   {We aim to measure the proper motion along the Sagittarius stream that is the missing piece to determine its full 6D phase space coordinates.}
   {We conduct  a  blind search of over-densities in  proper motion from the \Gaia second data release (DR2) in a broad region around the Sagittarius stream by applying wavelet transform techniques.  
   }
   {We find that for most of the sky patches, the highest intensity peaks delineate the path of the Sagittarius stream. The 1500 peaks identified depict a continuous sequence spanning almost $2\pi$ in the sky, only obscured when the stream crosses the Galactic disk. Altogether, around 100\,000 stars potentially belong to the stream as indicated by a coarse inspection of the colour-magnitude diagrams. From these stars, we determine the proper motion along the Sagittarius stream, making it the proper motion sequence with the largest span and continuity ever measured for a stream.
   A first comparison with existing N-body models of the stream reveals some discrepancies, especially near the pericentre of the trailing arm and an overestimation of the total proper motion for the leading arm.}
   {Our study can be the starting point for determining the variation of the population of stars along the stream, the distance to the stream with red clump stars, and the solar motion. It will also allow a much better measurement of the Milky Way potential.}


   \maketitle
%

\section{Introduction}\label{s:intro}

Research on tidal streams in the Milky Way has already profited immensely from the \Gaia mission \citep{Prusti2016}. The outstanding and precise proper motions of the second data release (DR2) \citep{Brown2018} fostered the discovery of many new streams both with known \citep{Ibata2019,Palau2019} and unknown \citep{Ibata2018,Malhan2018, Malhan2019} progenitors and the mapping over larger portions of the sky  \citep{Price-Whelan2018,Koposov2019}. \Gaia has also been crucial for the detection of peculiarities in streams such as diffuse components and gaps, which could give clues on the structure of the progenitor and of the Milky Way halo and on the dark subhalos that could have interacted with the stream stars \citep[e.g.][]{Bonaca2019,Malhan2018b,Malhan2019b,Bonaca2019b,Price-Whelan2018}. It has also revealed  evidence of misaligned proper motions, which seem to be the telltale of the gravitational effects of other Milky Way massive satellites \citep{Koposov2019,Erkal2019,Shipp2019}.

The stream of the Sagittarius dwarf (Sgr, \citealt{Ibata1994}), discovered by \citet{Mateo1996,Totten1998}, is the most prominent stream in the Milky Way halo and has served in multiple studies as a  prototype of ongoing tidal disruption and hierarchical formation of the Galaxy, and to constrain the Milky Way potential \citep[][and references therein]{Law2016}. 
Yet, reproducing the spatial structure simultaneously with the radial velocities remains a challenge. No single model has succeed in reproducing all observational data, with the best so far being the triaxial halo by \citet{Law2010} (hereafter LM10) and \citet{Deg2013}, giving a nearly oblate halo but with its minor axis oriented perpendicular to the Galactic plane. This configuration is infrequent in Lambda cold dark matter (LCDM) and is dynamically unstable, but might be avoided by considering the effect of the Large Magellanic Cloud \citep{Vera-Ciro2013} or a rising Milky Way rotation curve \citep{Ibata2013}. Even so, none of these models can account for other features of the Sgr stream like the bifurcations \citep{Belokurov2006,Koposov2012} and substructure near the apocentres \citep{Sesar2017}. The modelling efforts \citep{Fellhauer2006,Penarrubia2010,Dierickx2017} that have managed to reproduce these, however, still suffer from the radial velocity miss-match in the leading arm.

\begin{figure*}
   \centering

      \includegraphics[width=0.5\textwidth]{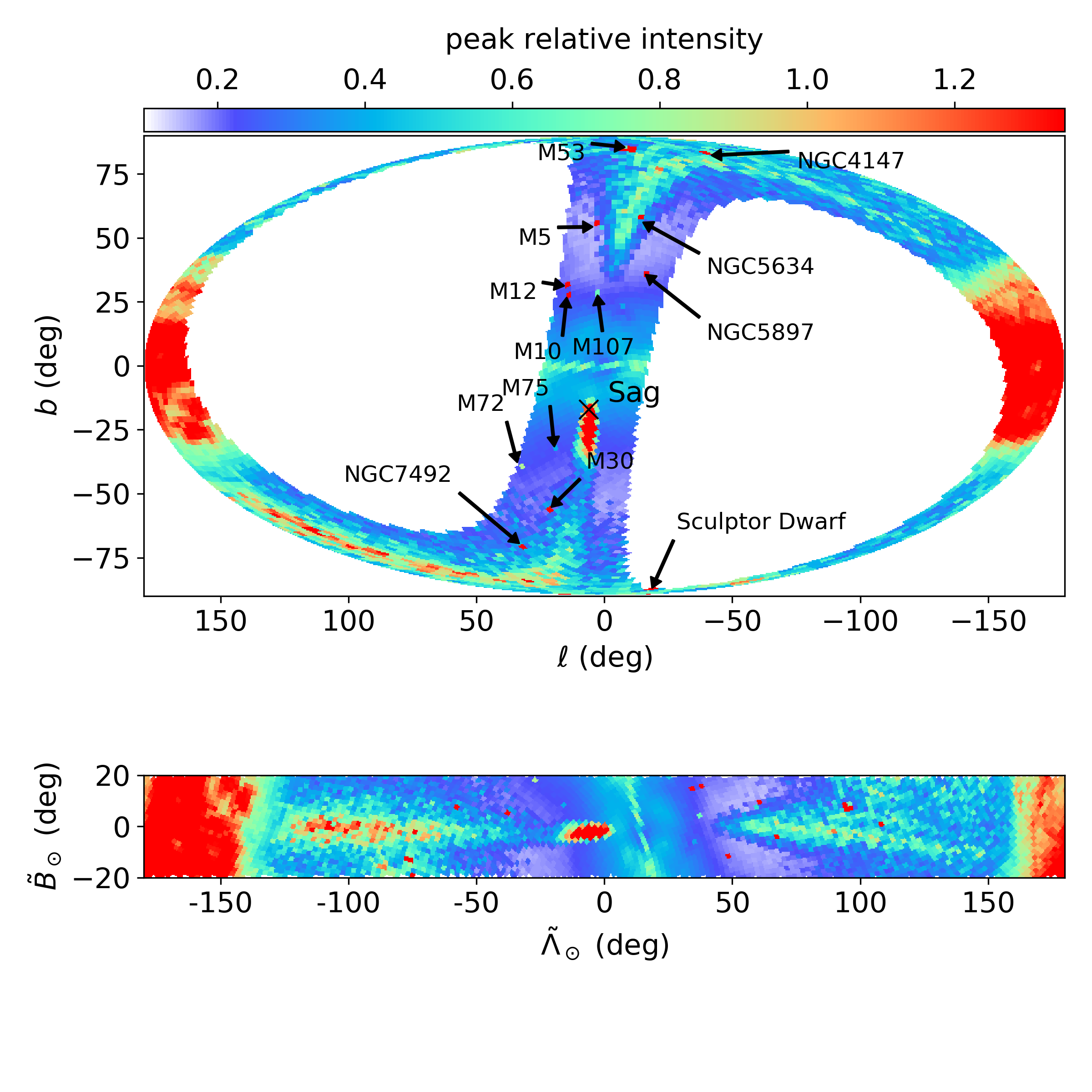}\hspace{-0.2cm}
   \includegraphics[width=0.5\textwidth]{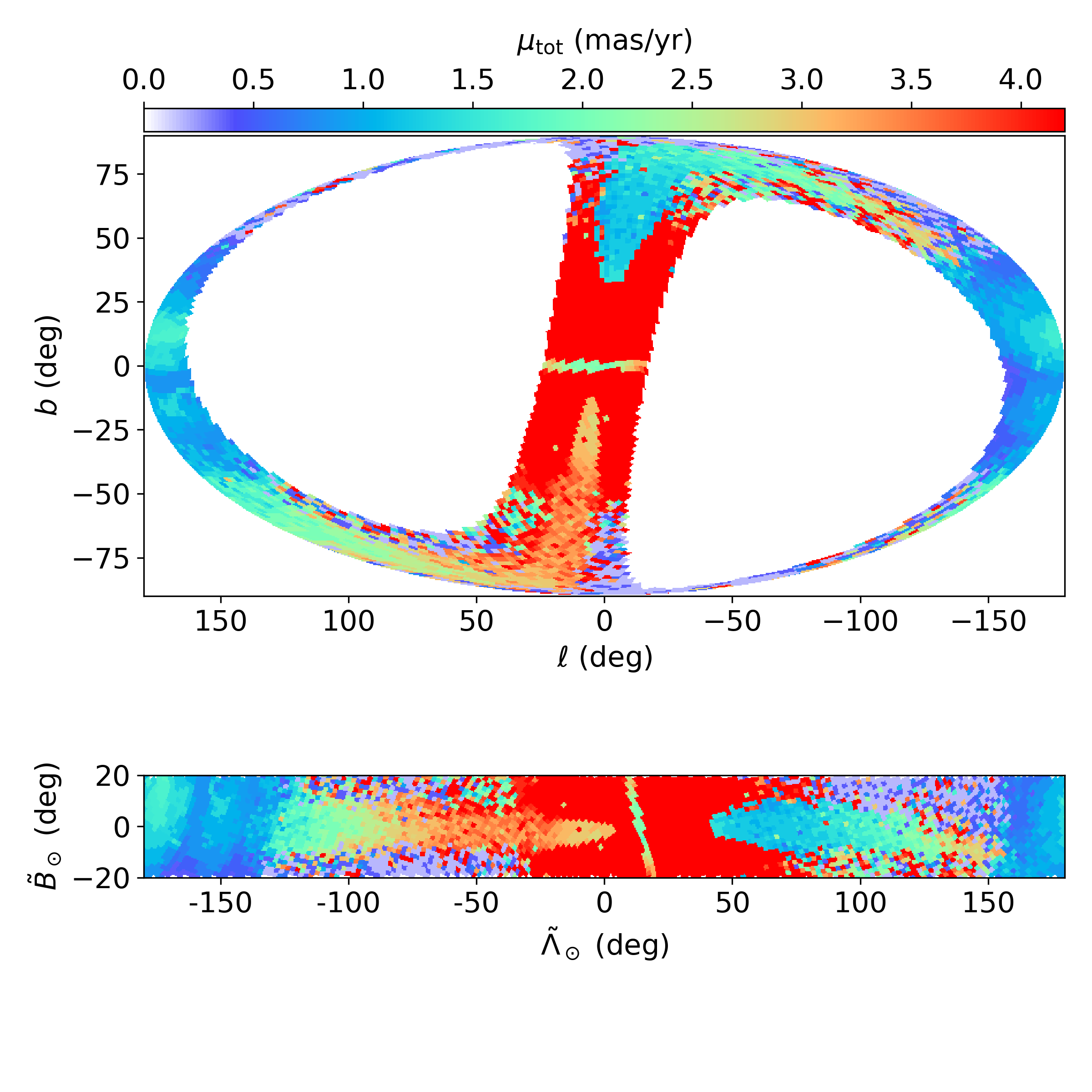}
 
     \caption{All sky view of the Sgr stream with Gaia data in Galactic coordinates ({\em top}) and in the coordinates of the orbital plane of Sgr ({\em bottom}). {\em Left:} \HP map in the region of the Sgr stream ($|\B|\leq20\deg$) coloured according to the relative intensity of the most prominent peak detected in the corresponding \HPf.  Known objects such as globular clusters and dwarf galaxies show high relative intensities in their corresponding \HP and are marked with arrows. The location of the Sgr dwarf is  shown with a black cross with position taken from \citealt{Karachentsev2004}. {\em Right:} Same as left but coloured as a function of the total proper motion $\mutot$. In all panels, the trace of the Sgr stream is clearly observed in an almost full circle on the sky, with the leading arm at positive Galactic latitude $b$ (positive $\L$) and the trailing at negative $b$ (negative $\L$).
     }
         \label{sky} 
   \end{figure*}  
   
A precise measurement of the proper motion along the Sgr stream could be the observational piece that is missing to solve these long-standing puzzles. So far, determinations of the Sgr proper motion have been very limited: they have
 only been obtained in small patches along the stream and are based on few selected member stars \citep{Carlin2012,Koposov2013,Sohn2015,Sohn2016}. Recently, \citet{Li2019} measured the proper motion of 164 stars from the Large Sky Area Multi-Object Fiber Spectroscopic Telescope (LAMOST) survey  \citep{Cui2012} along the trailing and leading arms with \Gaia DR2. \citet{Yang2019} detected substructure with a few hundreds of members related mostly to the leading arm of the stream, in a sample of LAMOST stars with proper motions from \Gaia DR2. 

Here we determine the proper motion along the Sgr stream using a completely different approach. We use {\it Gaia} DR2 astrometry and wavelet transform techniques to conduct a blind detection of over-densities in the proper motion plane for regions selected broadly in the plane of the stream (Sect.~\ref{s:datamethods}). 
We find, in an unforeseen way, that Sgr accounts for the peak of highest intensity in most of the fields probed (Sect.~\ref{s:sky}). From this we measure the continuous proper motion along the path of the stream (Sect.~\ref{s:pm}). By selecting all stars associated to the detected peaks, we explore their {\it Gaia} colour-magnitude diagram (CMD) and reconstruct the largest sample of candidate members of the Sgr stream (Sect.~\ref{s:population}). We discuss the many avenues opened by our findings in Sect.~\ref{s:conclusions}.

\section{Data and methods}\label{s:datamethods}

 We select stars from \Gaia DR2 \citep{Brown2018} with parallax satisfying $\varpi-\sigma_\varpi<0.1\,{\rm mas}$ and $BP-RP>0.2$. The first cut ensures that we keep stars farther than 10 kpc given their uncertainties and discards most of the foreground. The second cut eliminates blue foreground main sequence stars, while only removing a small fraction of blue
horizontal branch stars. 

We then look for over-densities in the proper motion covering the whole sky by applying a peak detection algorithm in the proper motion plane  $(\mua,\, \mud)$ of each \Gaia \HP of level 5. The peak detection algorithm is a simplified version of the method presented in \citet{Antoja2015b} based on the wavelet transform (WT) and it is detailed in Appendix~\ref{s:WT}.  
In summary, for the proper motion plane of each \HPf, the algorithm finds peaks and determines their proper motion coordinates, their significance, their width or scale, their intensity or height, and the number of stars that belong to the peak. The algorithm detects peaks of a set of discrete widths, named scales. Here we explore three logarithmically spaced scales in proper motion of 0.48, 0.96 and 1.92 $\masyr$, that we denote 1, 2 and 3. Finally, we select only the peaks with significance $\geq 3$ and we keep only the ones with highest relative intensity in each \HPf, that is, the peak with the largest height in each sky patch.

\section{A \Gaia full sky vision of the Sgr stream}\label{s:sky}

Figure~\ref{sky} shows two \HP maps of our results. We plot only \HP fields with $|\B|\leq20\deg$, where $\B$ is the latitude from the coordinate system of the Sgr stream $(\L,\B)$ defined in \citet{Majewski2003} with the orientation adopted in \citet{Belokurov2014}.  
In the left top panel each \HP is coloured as a function of the relative intensity of the highest peak. This map clearly shows the stream  emerging from the Sgr dwarf galaxy. The Sgr dwarf appears as a very strong concentration (red \HPf) at approximately the centre of the map, marked with a black cross. The leading arm ($b>0$) is seen in turquoise, while the trailing arm ($b<0$), having higher relative intensity, is seen in orange-red. This shows that in most of the sky around the Sgr stream the highest peak detected by our algorithm corresponds to the stream itself. 
Only where the stream crosses the Galactic plane, i.e. at $\ell\sim 0$ and $\sim 180\deg$, another dominant structure, probably related to the Milky Way disk, is apparent. We also see many other \HP with prominent peaks that correspond to known objects, mainly globular clusters and dwarf galaxies (some are indicated with black arrows). Their detection corroborates the 
reliability of our method. 

The \HP map in the top right panel of Fig.~\ref{sky} is coloured by total proper motion $\mutot$ of the highest intensity peaks. Again the Sgr stream appears as a river of \HP fields that gradually change their colour when moving from the Sgr dwarf towards the leading (blue-green colours) and trailing (orange-green colours) arms. 
This is the first all-sky view of the Sgr stream in proper motion ever produced.

In the bottom panels, the same \HP maps are shown but in the coordinates of the Sgr orbital plane as defined above. A clear bifurcation in the leading arm can be seen in the bottom left panel from $\L\approx90\deg$ to $\approx 140\deg$, confirming that the discovery by \cite{Belokurov2006} is observed also with \Gaia data. However, we do not see the bifurcation in the trailing arm \citep{Koposov2012} nor a signature that the bright and faint parts of the leading arm have different proper motion (bottom right panel).
\section{Proper motion sequence of the Sgr stream}\label{s:pm}
\begin{figure*}
\centering
    \includegraphics[height=0.28\textheight]{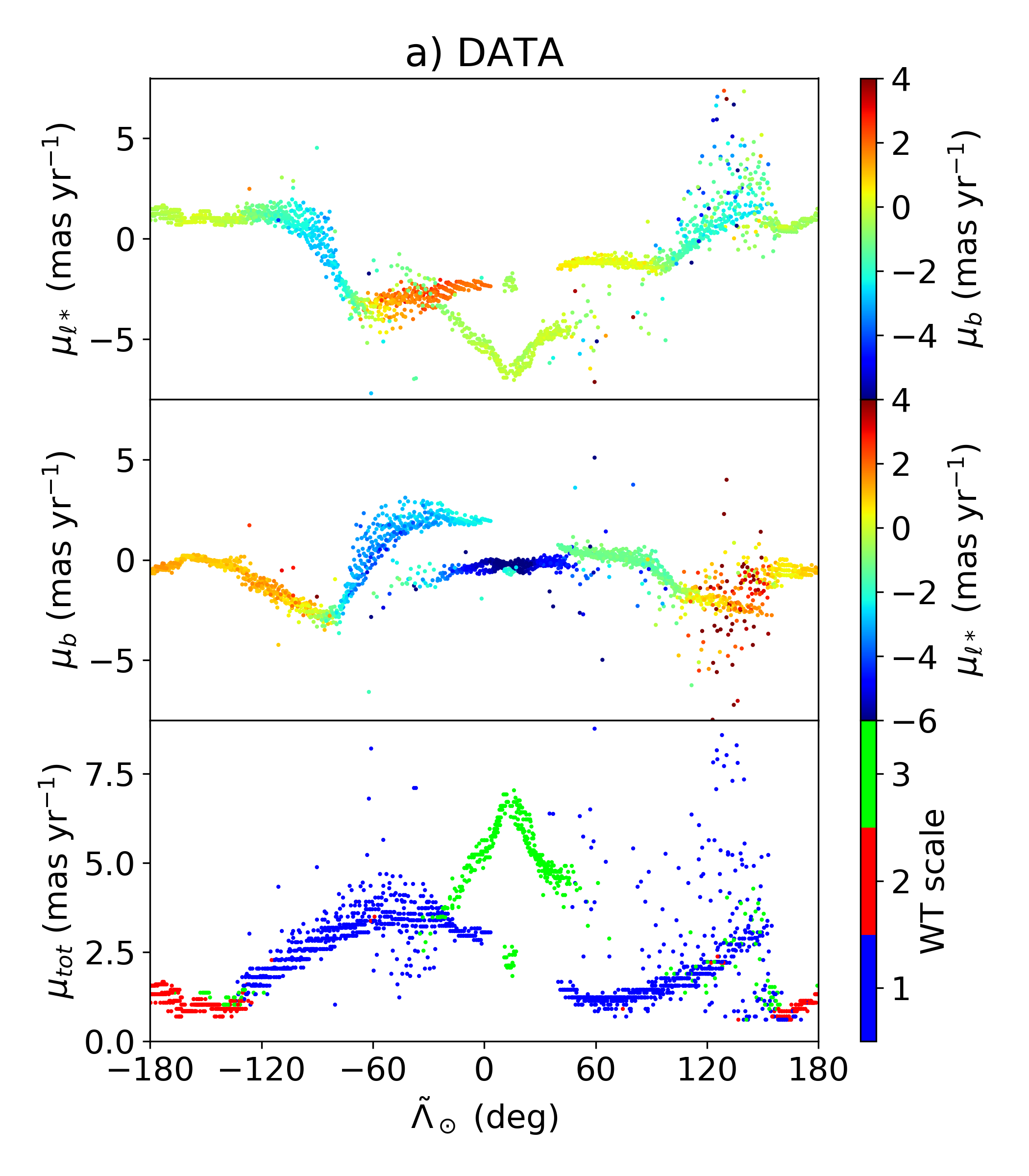}
    \includegraphics[height=0.28\textheight]{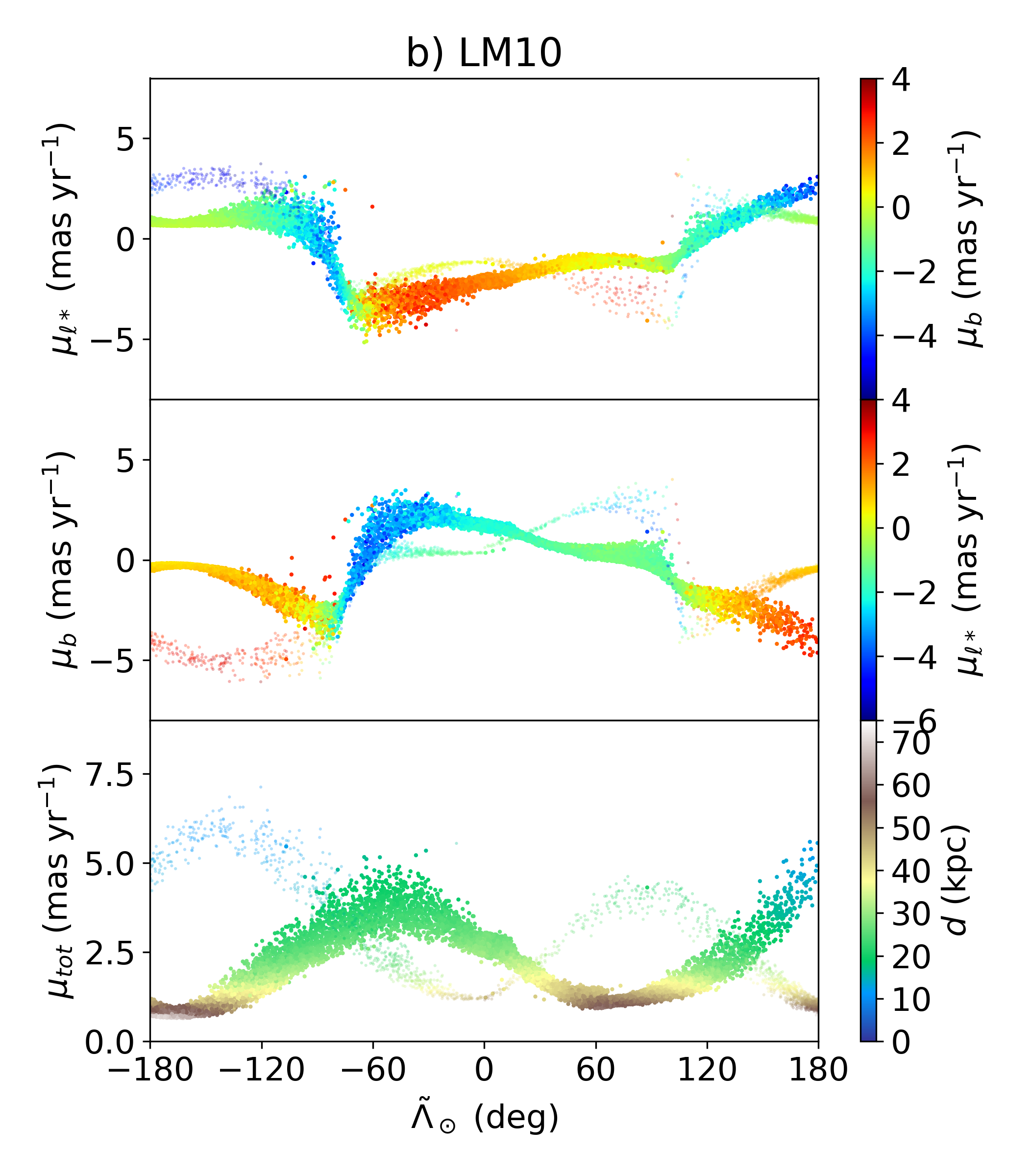}
        \includegraphics[height=0.28\textheight]{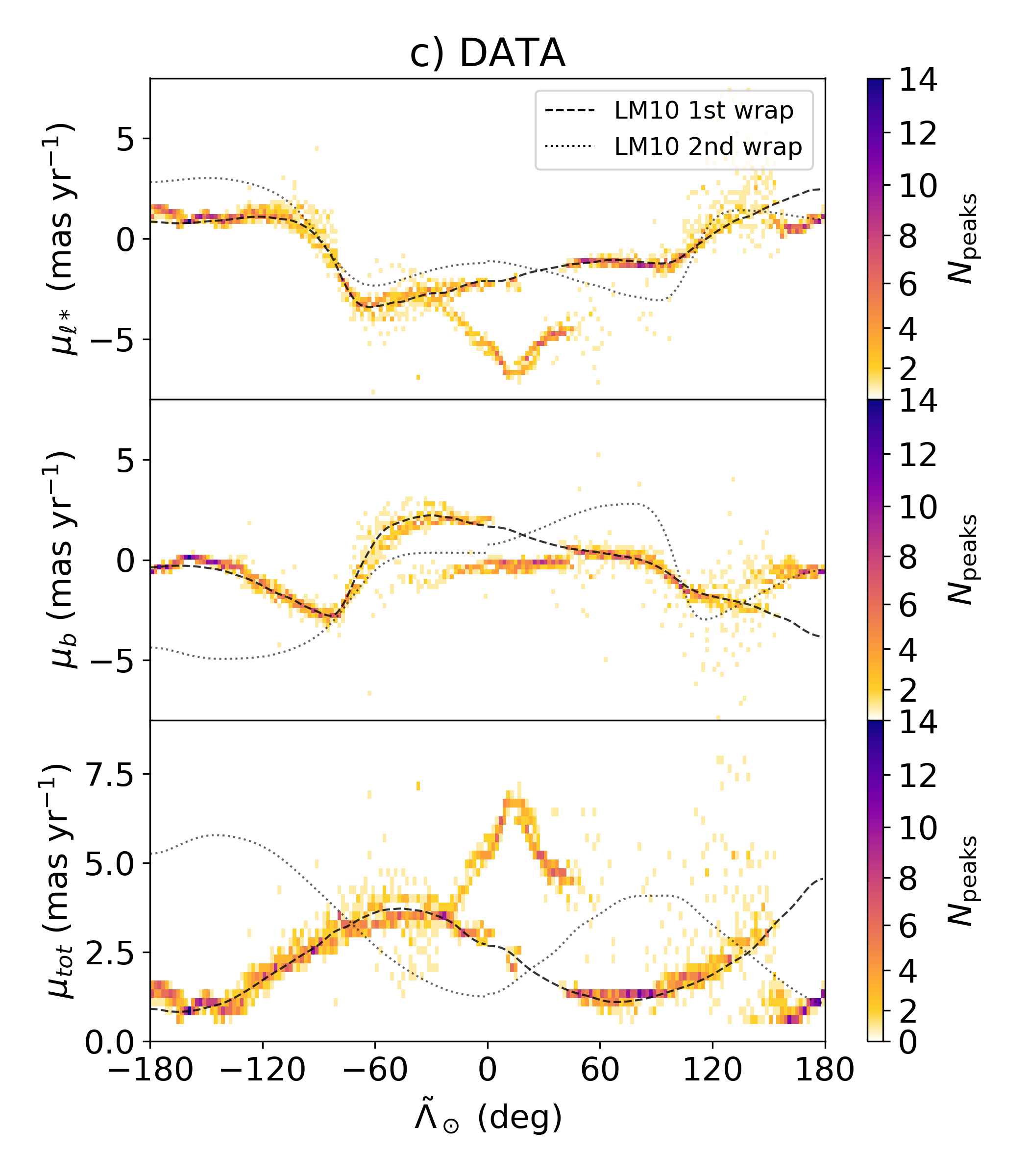}
      \caption{Proper motion of the Sgr stream from the \Gaia data and comparison with the LM10 model. All panels show the proper motion in Galactic longitude in the top, in latitude in the middle, and total in the bottom,  as a function of $\L$. a) Proper motions of the detected peaks colour coded by proper motion in latitude (top), in longitude (middle), in WT scale (bottom). b) Same but for the model and the bottom panel  coloured by distance to the Sun. The second wrap of the model has been given more transparency to enhance the contrast of the first wrap. c) Same as a) but represented in a two-dimensional histogram with the trace of the model superposed in black dashed (first wrap) and dotted (second wrap) lines.}
         \label{pm} 
   \end{figure*}
   
   \begin{figure*}
\centering
        \includegraphics[height=0.28\textheight]{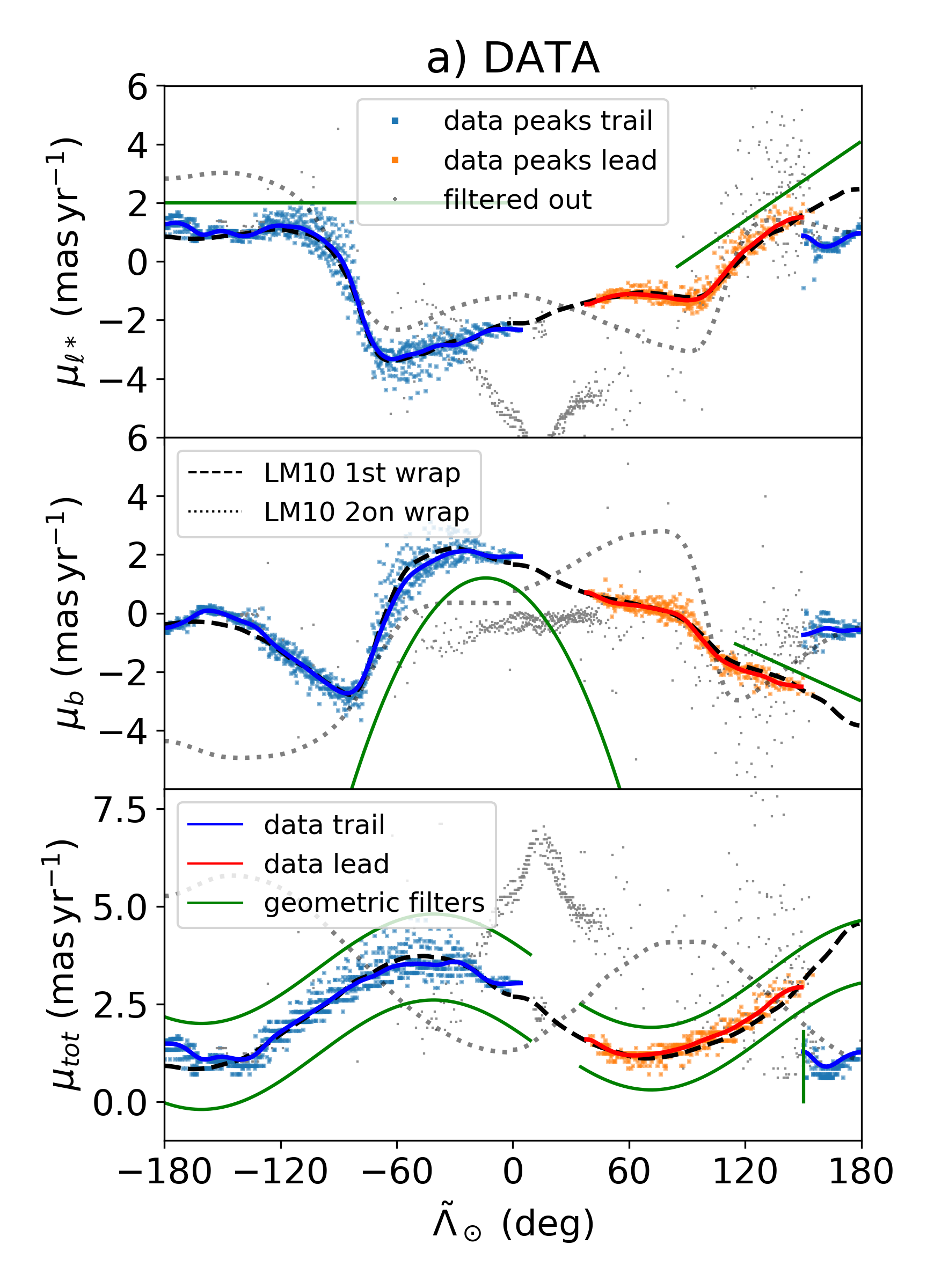}
         \includegraphics[height=0.28\textheight]{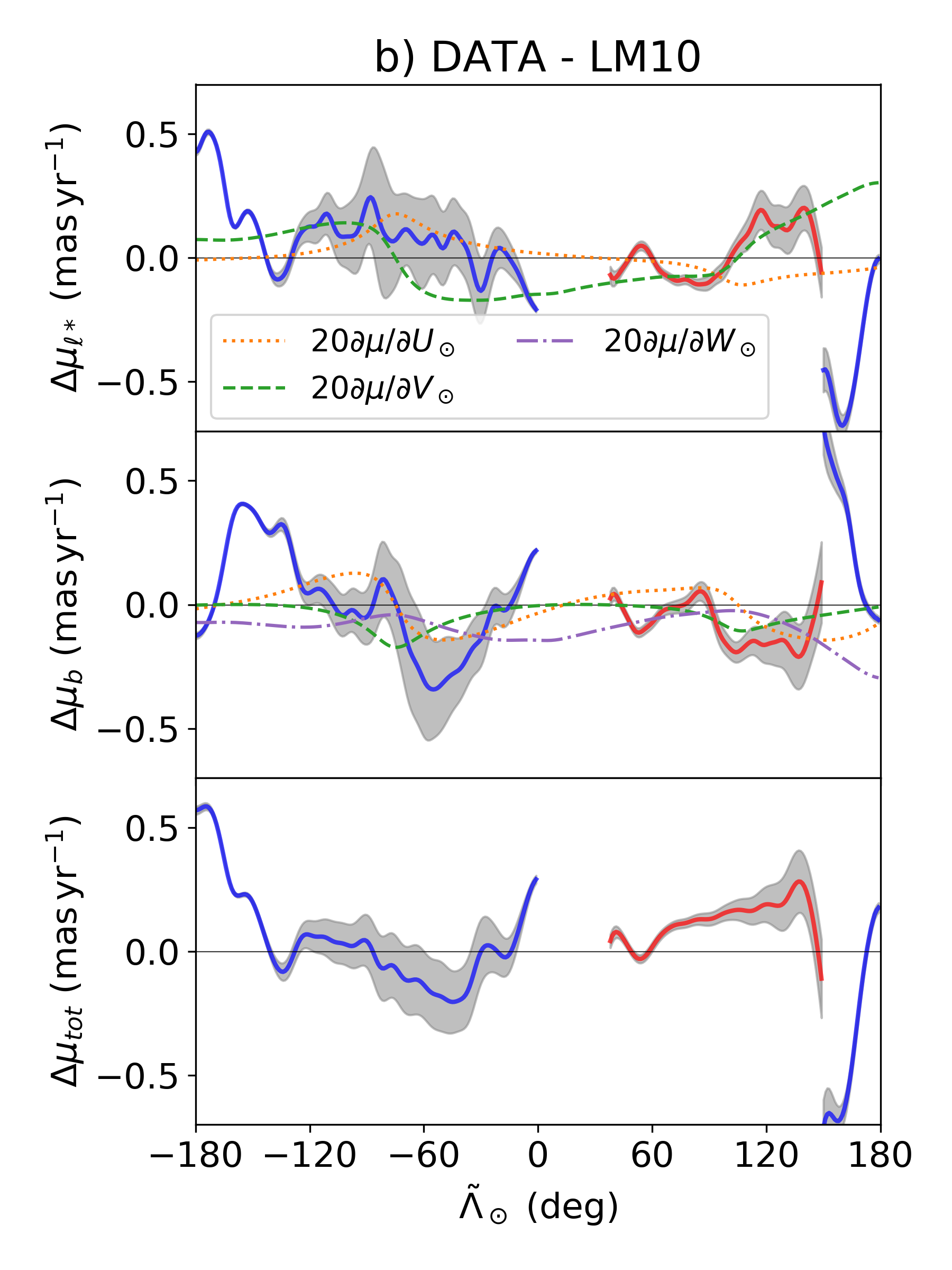}    
      \caption{Proper motion of the Sgr stream from the \Gaia data and comparison with the LM10 model (part 2). The top, middle, and bottom panels are organized as in Fig~\ref{pm}. a) Proper motions of the  peaks finally selected to belong to the Sgr stream coloured with blue (trailing) and orange (leading). A smooth version of the median proper motion of the \Gaia DR2 stars belonging to the detected peaks (see text) is shown in blue and red, together with the model in black. The green lines indicate the geometric filters that we apply to select the peaks that belong to the main sequences.  b) Differences in the median proper motions between data and model with the shaded areas indicating the statistical error. The rest of the curves indicate the values of a potential discrepancy due to an inappropriate assumption of the Solar motion values in LM10 by 20 $\kms$ (see text).}
         \label{pm2} 
   \end{figure*}
 
         
The proper motion of the peaks with the highest intensity are shown in Fig.~\ref{pm} as a function of $\L$, only for $|\B|\leq10\,\deg$ instead of $|\B|\leq20\,\deg$ as in Fig.~\ref{sky} in order to be more restrictive. Figure~\ref{pm}a shows the detected peaks in the planes of $\L$-$\mul$ (top), $\L$-$\mub$ (middle), $\L$-$\mutot$ (bottom) colour coded by $\mub$ (top), by $\mul$ (middle), and by the WT scale (bottom).
Figure~\ref{pm}b is the same as Figure~\ref{pm}a but for the LM10 model of the Sgr stream, in which we have selected only particles that roughly correspond to the observed parts of the stream\footnote{Particles in \url{http://faculty.virginia.edu/srm4n/Sgr/} with $P_{\rm col}\leq3$, i.e. that became unbound recently.}. Note that the data panels show the peaks detected in proper motion space (each one composed of hundreds to thousands of stars), while the panels for the model correspond to particles in the simulation. In the bottom panel of 
Figure~\ref{pm}b we colour code them by distance to the Sun. Figure~\ref{pm}c shows a 2D histogram of the data peaks. We overplot the LM10 model with black curves obtained by computing the median proper motion in bins of $\L$ and subsequently smoothing with a Gaussian filter. 
The data shows a continuous sequence of proper motion along $\L$ that strikingly resembles that of the LM10 model. We identify the first wrap of the trailing and leading arms of the Sgr stream. In addition, a few peaks seem to coincide with the expected position of the trailing arm at $\L\approx180\deg$. The continuity of the sequence is only broken where the stream crosses the Galactic disk and is hard to detect. 

In the data, there is a set of peaks at $\L\approx[-50,50]\deg$ organized along a triangular shape in the top and bottom panels that do not belong to Sgr. 
An analysis of a \Gaia mock catalogue without the Sgr stream (Appendix~\ref{s:mock}, Fig.~\ref{mock}) reveals a similar structure, hence, we conclude that
this is caused by contamination from the stellar foreground. We have filtered out these peaks by removing all peaks of scale 3 (green colours in the bottom panel of Fig.~\ref{pm}a). 
Additionally, we see peaks that fall outside the sequence around $\L\approx150\deg$. These do not appear in the mock catalogue but appear in an off-stream sky band selected in the range $15<|\B|\leq 20\deg$ (Appendix~\ref{s:mock}, Fig.~\ref{off}). Their origin at this point is not clear but we note that their location in the sky is similar to that of the Virgo over-density and there is some overlap in proper motion according to the values reported by \citet{Yang2019}.

For a more quantitative comparison between the data and the model, we refine the selection of peaks by applying simple geometric filters in the three planes ($\L$-$\mul$, $\L$-$\mub$, $\L$-$\mutot$), intending only to remove peaks that are clearly off the proper motion sequence. These filters are detailed in Appendix~\ref{s:selection} and the region where peaks are selected is shown with green lines in Fig.~\ref{pm2}a. We finally end up with 1540 peaks which are shown in blue and orange for the trailing and leading arms, respectively. Then we retrieve from the {\it Gaia} Archive the stars associated to each peak by selecting those inside a circle centred on each peak, with a radius equal to the corresponding WT scale. From this set of stars we have removed stars from globular clusters as detailed in Appendix~\ref{s:globular}. Our final sample has 2\,168\,723
 stars. 

We then compute the stars median $\mul$, $\mub$ and $\mutot$ in bins of $\L$, which we then smooth with a Gaussian filter, as done previously for the model. We compute the statistical error on the median using the approximation $\sqrt{\frac{\pi}{2}}\frac{\sigma}{N}$, where $\sigma/N$ is the standard error on the mean. The curve obtained is shown in Fig.~\ref{pm2}a with blue (trailing) and red (leading) solid lines. The error is 
 $<$0.03$\,\masyr$ for 75\% 
of the points in the smoothed curves. Again, the black lines correspond to the LM10 model.

Finally, Fig.~\ref{pm2}b shows the  $\Delta\mul$,  $\Delta\mub$ and  $\Delta\mutot$ residuals, 
defined as $\Delta \equiv ({\rm data})-({\rm model})$
with the grey area showing the combined errors on the median added in quadrature (thus, $1\sigma$). The differences are  $< 0.2\,\masyr$ for 75\% of the bins along the evaluated sequences. For the leading arm, we see systematic differences $\lessapprox 0.2\,\masyr$ in $\mul$ and $\mub$ well above the statistical error and a larger total proper motion than for the model. The biggest differences, though, are for the trailing arm at $\L<-120\deg$ (including its prolongation at $\L\approx180\deg$). These discrepancies could be due to a strong contamination from the foreground (see Appendix~\ref{s:mock}). Indeed, the CMDs analysed in Sect.~\ref{s:population} appear very contaminated in this range. 
There are also significant discrepancies near $\L\approx-50 \deg$ 
in 
$\mub$ that 
 are $\sim0.3\,\masyr$. 
This is close to pericenter of the orbit (small distances, see bottom panel of Fig.~\ref{pm}b). 
This position is also where the contribution of the Solar motion is larger. In Fig.~\ref{pm2}b we plot the derivative of the contribution of Solar motion with respect to the individual components $(U_\odot$, $V_\odot$, $W_\odot)$. A distance needs to be assumed for this calculation, so we take it from the LM10 model. We see that the shape of $\frac{\partial \mul}{\partial U_\odot}$ and $\frac{\partial \mub}{\partial U_\odot}$ is similar to that of the differences between data and model for the trailing arm, but not for the leading arm. However, all the derivatives are very small (at most of $0.01\,\masyr\,\kms$). Hence, the Solar motion assumed in the LM10 would have to differ by more than $20\,\kms$ in order to explain the observed differences. This could be plausible for $V_\odot$ (for instance $V_\odot=232\,\kms$ in LM10 but $V_\odot=255\,\kms$ from \citealt{Reid2014}) and, therefore, at this point we cannot discard that a possible combination of a different $V_\odot$ together with a distance difference is the cause of part (but not all) of the proper motion discrepancies. 

The systematic errors in the \Gaia proper motions cannot account for  the observed residuals since they are much smaller (of the order of $0.028\,\masyr$ at the scales of $\sim14$–$20\deg$, see Table 4 in \citealt{Lindegren2018}). 

\section{300\,000 stars in the Sgr stream and Sgr dwarf}\label{s:population}

  \begin{figure*}
   \centering
      \includegraphics[width=1.\textwidth]{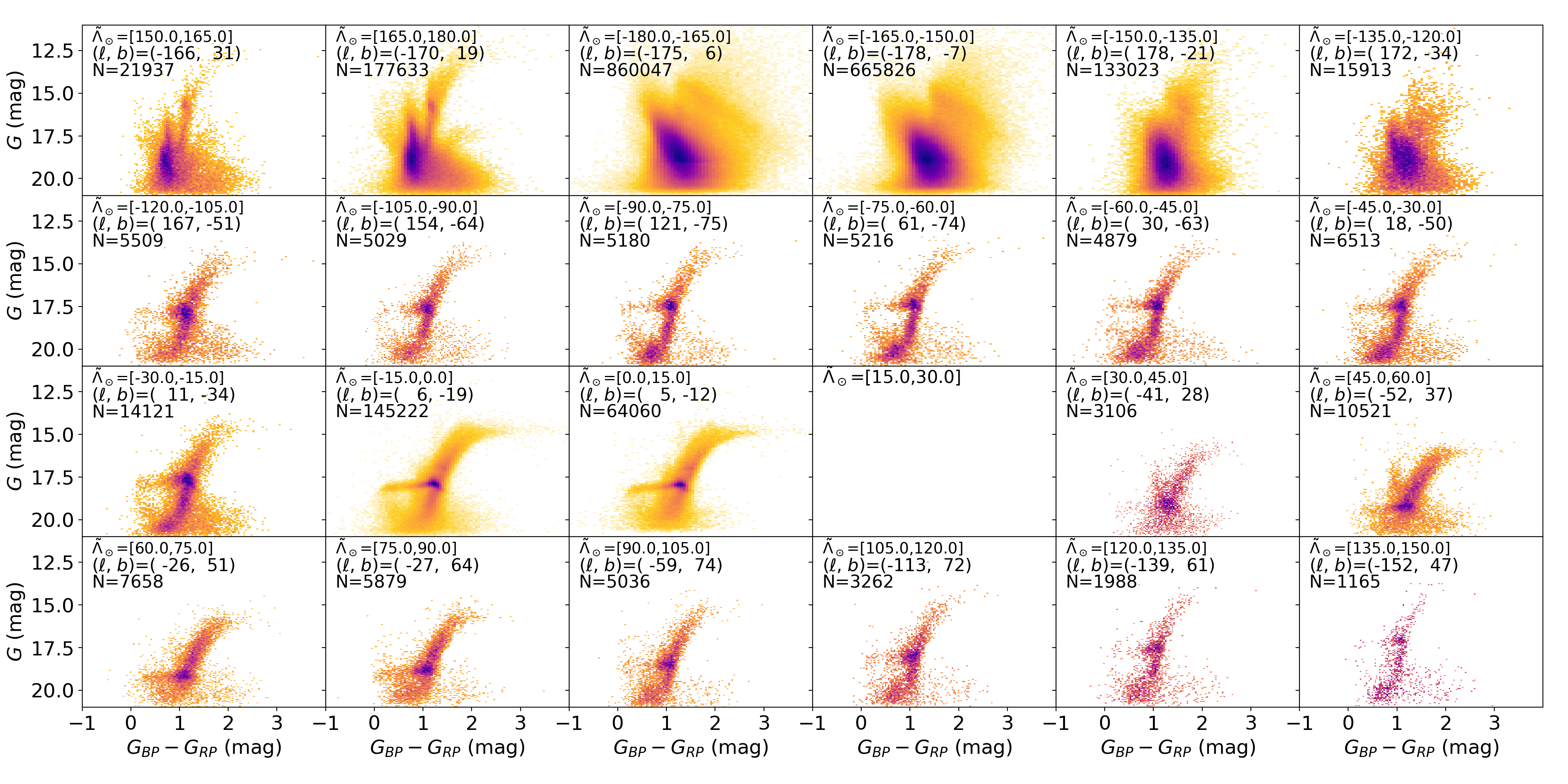}

      \caption{Colour-magnitude diagrams of the stars belonging to the proper motion peaks classified as Sgr peaks for different bins in $\L$ as indicated in the legend. The corresponding means in Galactic coordinates of the stars are also shown. The CMDs at higher Galactic latitudes are dominated by the Sgr population.}
         \label{cmd} 
   \end{figure*}
   


In Fig.~\ref{cmd} we plot the CMDs of the stars that belong to the detected peaks of Sgr in  bins of $15\deg$ in $\L$. We have organized the panels so that the detection of the trailing arm at $\L=180\deg$ is connected to that at $\L=-180\deg$. In the direction of the Galactic centre, there were no selected peaks and hence the blank panel. Except for the panels of the first row that are close to the Galactic plane in the anti-centre direction (see discussion below), the  signature of Sgr is clear, with prominent red clump (RC), red giant branch, extended horizontal branch, and turn-off stars. The Sgr dwarf itself has high density of stars in the range $\L=[-15,15]\deg$ (second and third panels of the third row).

 \begin{figure}
   \centering
      \includegraphics[width=0.5\textwidth]{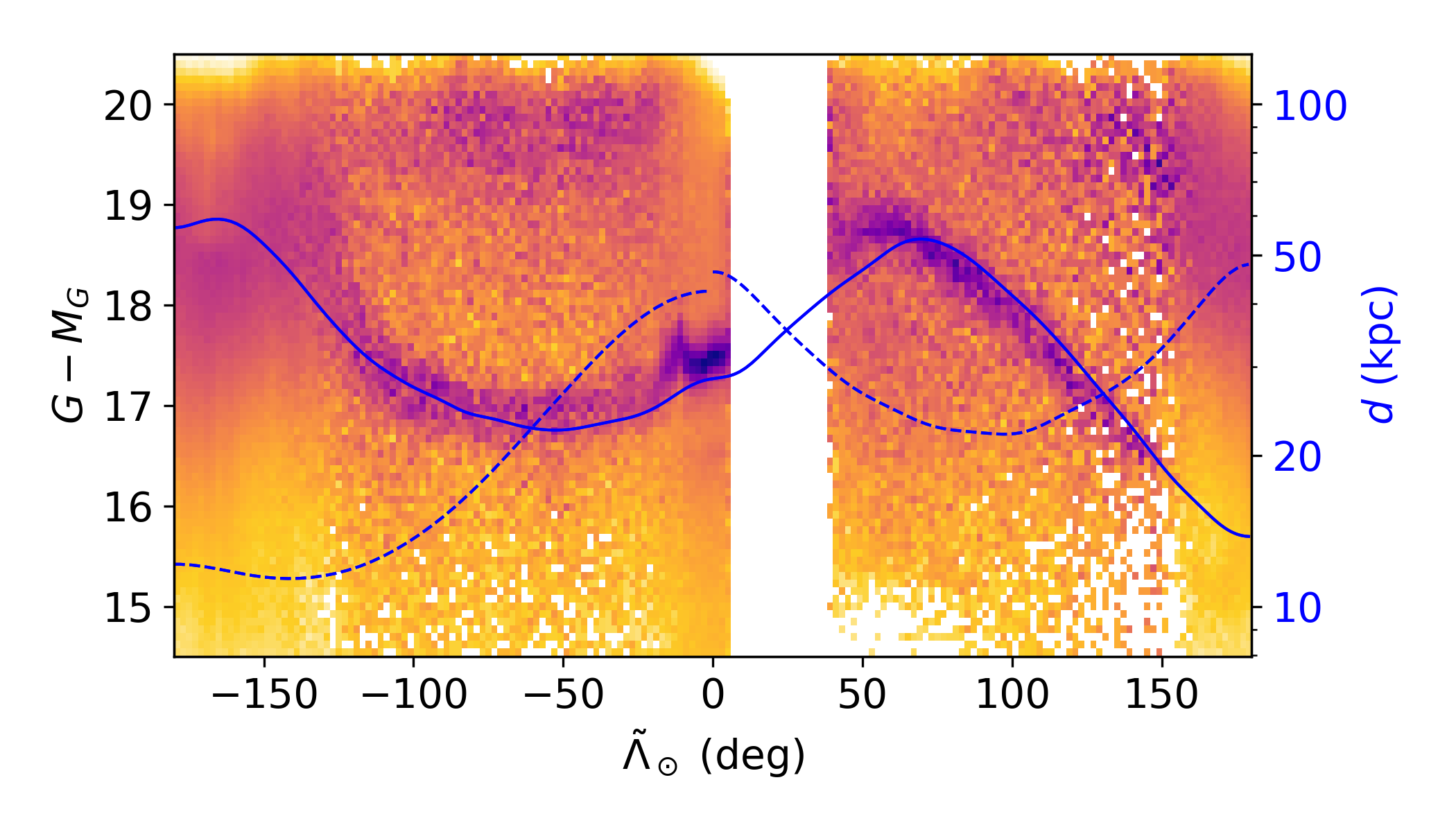}
      \caption{Apparent distance modulus as a function of $\L$ for the stars belonging to the Sgr stream and Sgr dwarf. The data is shown as a histogram normalized by column. The stars belonging to the RC depict a  well defined sequence. In a secondary axis, the distance of the particles in the LM10 model are superposed in blue. In absence of extinction ang with our assumed calibration of the RC, the two would be directly comparable. The data shows general good agreement with the model with discrepancies e.g. in the apocenter of the leading arm.}
         \label{rc} 
   \end{figure}
   
In Fig.~\ref{cmd} we observe the position of the RC moving  continuously in $G$ magnitude as a function of $\L$, due to the varying distance of the stream. 
This is better seen in Fig.~\ref{rc} where we show the apparent distance module versus $\L$, computed as if all stars were RC stars, assuming a RC absolute magnitude of $M_G=0.495$ \citep{RuizDern2018} and neglecting effects of extinction and of the metallicity gradient along the stream \citep[e.g.][]{Keller2010,Hayes2019}. 
The distance trend shows general good agreement with the LM10 model (blue curves, right axis). There are  differences of the order of 5 kpc in the leading arm at $\L\approx100\deg$, which go in opposite sense of a potential extinction effect, and a shift in the angular position of the apocentre at around $\L\approx70\deg$. A more thorough comparison of the distances of data and model will be a subject of future work (Ramos et al., in prep).

We note that a wrong distance in the LM10 model would also cause a discrepancy in the proper motions. 
For instance, an overestimation of the model distance for the leading arm by $10\%$ will cause an overall underestimation of the proper motion of $10\%$, which is actually similar to what we observe. 
We do not observe, however, a difference in the distances that could account for the strong proper motion discrepancies and their sudden sign changes in the trailing part.

It is hard to identify the Sgr trailing stream in the CMDs of the first row of  Fig.~\ref{cmd} 
 where there is strong contamination. The RC of the trailing arm should be quite faint here given the distances of this part of the stream ($\sim 50$ kpc according to LM10) and is probably hidden inside the big blob of contaminating stars. We only see the emerging tip of the giant branch at $\L=[150,180]\deg$ (plume of stars at $G_{BP}-G_{RP}\approx1.75$) and some hints of it at $\L=[-150,-120]\deg$. The RCs we see in the first five panels of Fig.~\ref{cmd} have a magnitude that could be compatible with the distance to the leading arm in its second wrap but the proper motion of our selected peaks makes it very unlikely\footnote{Indeed, the proper motion difference of the two branches is about 5 $\masyr$ according to the LM10 model, which is much larger than the expected dispersion of the stream considering the intrinsic dispersion and also the one produced by errors.}. The structure that we are seeing in these panels could be then due to contamination of the disk/s, the flare, the warp or other outer disk substructures \citep{Xu2015} such as the Monoceros ring \citep{Newberg2002} and TriAnd overdensity \citep{Rocha-Pinto2004,Sheffield2014,Price-Whelan2015}.

 We note that our selection of member stars is contaminated in certain areas and that the completeness resulting from our methodology is yet to be determined. While we plan to deal with these in the future, here 
we build a list of member \emph{candidates} by selecting stars in the ranges of $\L \in [-120,\,-10]\deg$ (trailing arm) and of $\L\in [ 30,\,150]\deg$ (leading arm) where the contamination is minimal. We count 
 61\,937 and 
38\,615 stars, in the trailing and leading arms, respectively, making a total of 
 100\,552 
stars of different stellar types along the stream of Sgr. 
 We also find 193\,792 stars in the approximate region of the Sgr dwarf ($\L=[-10,10]\deg$), given an apparent size of 15 $\deg$ \citep[e.g.][]{Giuffrida2010}. 
 
 The list of member candidates (see Table~\ref{table:1}), the median proper motions of the data, the interpolation used to obtain the smooth curves given as a pickle \textit{Python} object, and additional material will be made publicly available at \href{https://services.fqa.ub.edu/sagittarius}{https://services.fqa.ub.edu/sagittarius} and through CDS.

\section{Discussion and conclusions}\label{s:conclusions}

We have presented a measurement of the proper motion of the Sgr stream with {\it Gaia} DR2 data. Compared to previous work based on few stars located in a few particular fields \citep{Carlin2013,Koposov2013,Sohn2015,Sohn2016}, our determination is substantially better since: i) it is based on much larger number of stars (around 100\,000 stars in the stream) and covers different stellar types, ii) it is, for the first time, a continuous determination that extends, except for the regions behind the Galactic plane,  $2\pi$ in the sky. This makes it the proper motion determination with the largest span for any stream to date.  We also detect clearly the bifurcation of the leading arm. The whole sky coverage and precision of the proper motions given by \Gaia are the key to the success of the present study.

Our determination comes directly from the detection of peaks of stars in the proper motion plane covering the sky continuously. That is, we do not use a previous selection of Sgr members, but the member candidates are rather a by-product of our study. In this sense, it is the first time that the stream is detected based on a proper motion search. Moreover, our search is a blind one, meaning that the peaks corresponding to the over-density of the stream stand out naturally as the peaks with highest intensity in the proper motion distribution of each sky patch. We note that such a continuously determination would be hard to achieve using match filtering techniques since the distance to the stream changes across the sky. Moreover, our method does not need any assumptions contrary to other methods that use also proper motion information for the stream detection such as in STREAMFINDER \citep{Malhan2018}
where they look for trajectories coherent with a stream shape in a given gravitational potential. In fact, in the final stages of this work we found out that a simultaneous study by Ibata et al. (private communication) also detected the Sagittarius stream with Gaia data using this completely different methodology and it will be extremely interesting to compare their results with ours in the near future. We note, however, that our ability to detect new streams as done with STREAMFINDER has still to be proven.

We note that we detected several globular clusters along the path of the Sgr stream that also fall in the same proper motions sequence as Sgr. From these, there is consensus about the likely relation of Pal\,12, Arp\,2, M\,54, Ter\,7 and Ter\,8 with Sgr \citep[e.g.][]{Law2010b,Massari2019}, and more recently of NGC\,2419  \citep{Belokurov2014b,Massari2017,Sohn2018} that we confirm here, while not for NGC\,5053 and NGC\,5024   \citep{Sohn2018,Tang2018}.

Our list of member candidates contains about 100\,000 stars in the stream and 200\,000 stars in the Sgr dwarf itself and 
opens up many possibilities for further investigations, 
such as on the population of stars along the stream, the distance to the stream using RC stars, the width of the stream in the sky and its velocity dispersion, 
 the determination of the solar motion \citep{Hayes2018}, the possible detection 
 gaps  and spurs 
 which are expected \citep{Amorisco2015} and seen in other streams \citep{Price-Whelan2018,Erkal2019}. 
The continuity and span in the sky of our proper motion measurement allows for a determination of the 3D kinematics of the Sgr stream along its first $2\pi\deg$ that could hold the clue for an improvement of the models of the gravitational potential of the Galaxy and companions such as the LMC influencing the Sgr orbit \citep{Vera-Ciro2013}.

\begin{acknowledgements}
 We thank the anonymous referee for his/her helpful comments. This work has made use of data from the European Space Agency (ESA) mission {\it Gaia} (\url{https://www.cosmos.esa.int/gaia}), processed by the {\it Gaia} Data Processing and Analysis Consortium (DPAC, \url{https://www.cosmos.esa.int/web/gaia/dpac/consortium}). Funding for the DPAC has been provided by national institutions, in particular the institutions participating in the {\it Gaia} Multilateral Agreement. 
This project has received funding from the European Union's Horizon 2020 research and innovation programme under the Marie Sk{\l}odowska-Curie grant agreement No. 745617 and No. 800502.
This work was supported by the MINECO (Spanish Ministry of Economy) through grant ESP2016-80079-C2-1-R and RTI2018-095076-B-C21 (MINECO/FEDER, UE), and MDM-2014-0369 of ICCUB (Unidad de Excelencia 'Mar\'\i a de Maeztu').
 This project has received support from the DGAPA/UNAM PAPIIT program grant IG100319.
AH acknowledges financial support from a VICI grant from the Netherlands Organisation for Scientific Research, NWO. JAC-B acknowledges financial support to CAS-CONICYT 17003. CM is grateful for the hospitality of the ICCUB, during visits in which part of this research was carried out.  
\end{acknowledgements}


\bibliographystyle{aa}
\bibliography{mybib}

\begin{appendix}

\section{Peak detection algorithm and query in the \Gaia Archive}\label{s:WT}
 
 The peak detection algorithm is a simplified version of the method presented in \citet{Antoja2015b}. Here we summarize the algorithm but more technical details and recent applications can be found in \citet{Starck2002,Antoja2015b,Ramos2018}. 

\begin{figure}
   \centering
      \includegraphics[width=0.48\textwidth]{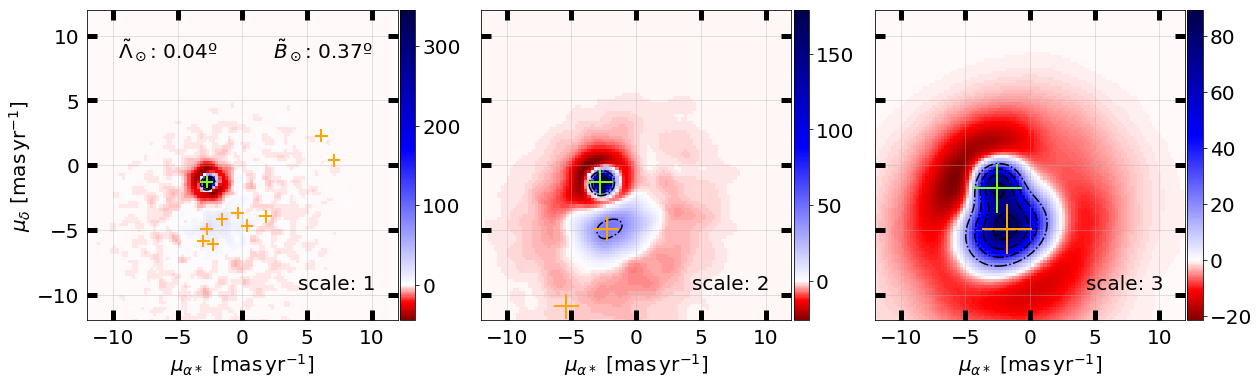}

      \includegraphics[width=0.48\textwidth]{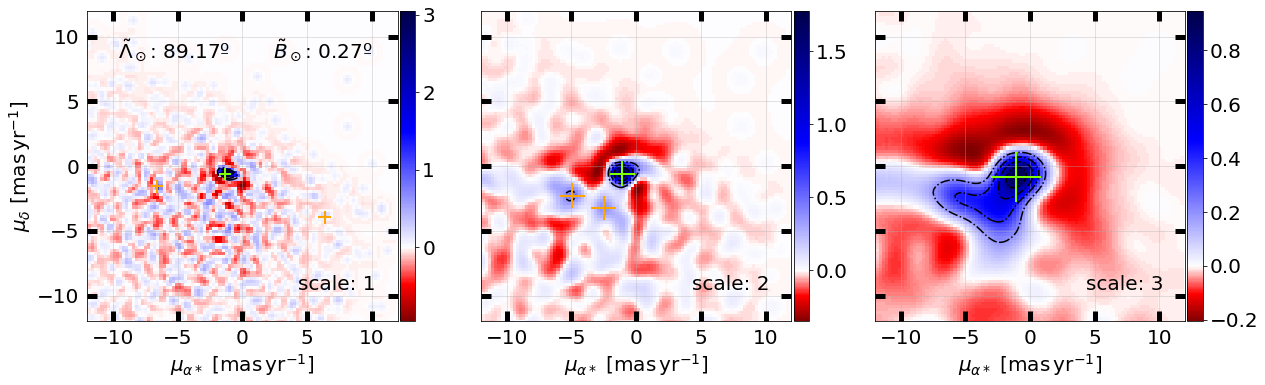}

      \caption{Example of WT planes in two different \HP fields corresponding to the Sgr core (top row) and the middle part of the leading arm (bottom) at the three different scales analyzed here (three columns, from left to right for small to large). The colour indicates the value of the WT coefficients, where darker blue means higher WT, i.e. higher intensity of the substructure, while red colours are for under-dense regions. Crosses are peaks detected in these WT planes.  The peak of the Sgr stream/dwarf is marked in green and appears well separated from the general field.}
         \label{pmplane} 
   \end{figure}

The main piece of the algorithm is the {\em a trous} (‘with holes’) variant of the wavelet transform (WT, \citealt{Starck2002}), which computes a discrete set of scale-related ‘views’ of a 2D image, where substructure at a particular scale is highlighted. We use the implementation in the MultiResolution software\footnote{\url{http://www.multiresolutions.com/mr/}}. For each \HPf, our initial image is the 2D histogram in proper motion space with bin size of 0.24 $\masyr$ and range of $[-12, 12]$ $\masyr$. We explore a set of three logarithmically spaced scales in proper motion of 0.48, 0.96 and 1.92 $\masyr$, that we denote 1, 2 and 3. An example of the WT application is presented in Fig.~\ref{pmplane}, where the WT coefficients are shown in a blue (red) colour scale corresponding to over-dense (under-dense) regions, for the three different scales (columns) and for two different sky positions as explained in the labels (rows).

Then we search for local peaks in the WT plane at each scale, using {\tt{peak\_local\_max}} from the \textit{Python} package {\tt{ Scikit-image}} \citep{vanderWalt2014}. The WT also gives the significance of the coefficients based on a model for the Poisson noise. From all the peaks detected, we select only those with WT coefficient with significance\footnote{This is equivalent to $WP\geq99.7$ in the notation of \citet{Antoja2015b} and to $CL=3$ in \citet{Ramos2018}.} $\geq3$, which corresponds to $\sigma\geq3$ in the Gaussian-equivalent case. The peaks detected in each scale are marked with crosses in Fig.~\ref{pmplane}. In these plots, which correspond to a \HP centred close to the Sgr dwarf (top) and the leading arm (bottom), respectively, a prominent peak is found (green cross) that corresponds to Sgr, as we shall see in next sections.
 
Finally, the method yields the list of significant peaks in proper motion at each \HP with their associated quantities such as the location of the peak in  $\mua$ and $\mud$, the wavelet transform coefficient WT, which can be thought of as the intensity of the peak, the scale where it was detected (1, 2 or 3), and the number of stars belonging to the peak (computed as described in \citealt{Ramos2018}). 
A measure of the relative intensity of a peak can be computed as:
\begin{equation}\label{RI}
    \frac{WT}{N_{\mathrm{hp}}}\times 1000,
\end{equation}
where $N_{\mathrm{hp}}$ is the total number of stars in the \HP and the 1000 factor is used to convert to more usable numbers (typical relative peak intensities will be of the order of 1). 

The peak associated to the Sgr stream is often detected at more than one scale in each \HP as exemplified in Fig.~\ref{pmplane}. However, the peak will have highest relative intensity at the scale that is closer to its actual size (typically, as in Fig.~\ref{pmplane}, this occurs at scale 1, that is 0.48 $\masyr$).
We will select from now on only the peaks with the highest relative intensity for each \HPf, which will naturally remove redundant peaks.

Due to the impracticality of downloading all the stars that fulfill the cuts imposed, we process in parallel each HEALpix downloading from the \Gaia Archive directly the proper motion histograms using:

\begin{tt} SELECT COUNT(*) as N, pmra\_index*\textit{binsize} as pmra, pmdec\_index*\textit{binsize} as pmdec FROM (SELECT source\_id, FLOOR(pmra/\textit{binsize}) AS pmra\_index, FLOOR(pmdec/\textit{binsize}) AS pmdec\_index FROM gaiadr2.gaia\_source WHERE source\_id BETWEEN  \textit{hpnum}*2**35*4**(12-\textit{level})  AND  (\textit{hpnum}+1)*2**35*4**(12-\textit{level}) AND parallax-parallax\_error < 0.1 AND bp\_rp >= 0.2 AND pmra IS NOT Null AND pmdec IS NOT NULL) as sub GROUP BY pmra\_index, pmdec\_index \end{tt}

\noindent where one should replace \textit{hpnum}, \textit{level} and \textit{binsize} by the number of the HEALpix, the level of the HEALpix grid (here 5) used, and the size of the bin for the histogram (here 0.24 $\masyr$).
 This results in fast queries that return easy to handle small size files. Then we apply the WT and peak search algorithm. With a regular computer and 8 CPU cores running, the whole sky execution takes of the order of 12h. 

In several \HPf we found peaks at ($\mua$, $\mud$)$\approx (0,0)$. These are triggered by quasars: in several regions of the sky, especially where the \Gaia coverage has been larger (e.g. large {\tt visibility\_periods\_used}), 
quasars are observed more easily and appear as concentrations of sources with almost zero proper motion. 
Those peaks have been removed by selecting only peaks with $\mutot\equiv\sqrt{\mua^2+\mud^2}>0.6\,\masyr$. Although this could be hindering the detection of the parts of the stream with small proper motions, most of these occurs in areas that appear to be strongly contaminated by the foreground and would not have entered our selection (see Sect.~\ref{s:pm}). We also note that the LM10 model predicts very few stars ($<2\%$) with $\mutot<0.6\,\masyr$. 



\section{Tests with mock \Gaia DR2 data and fields off-stream}\label{s:mock}

\begin{figure}
   \centering
         \includegraphics[width=0.24\textwidth]{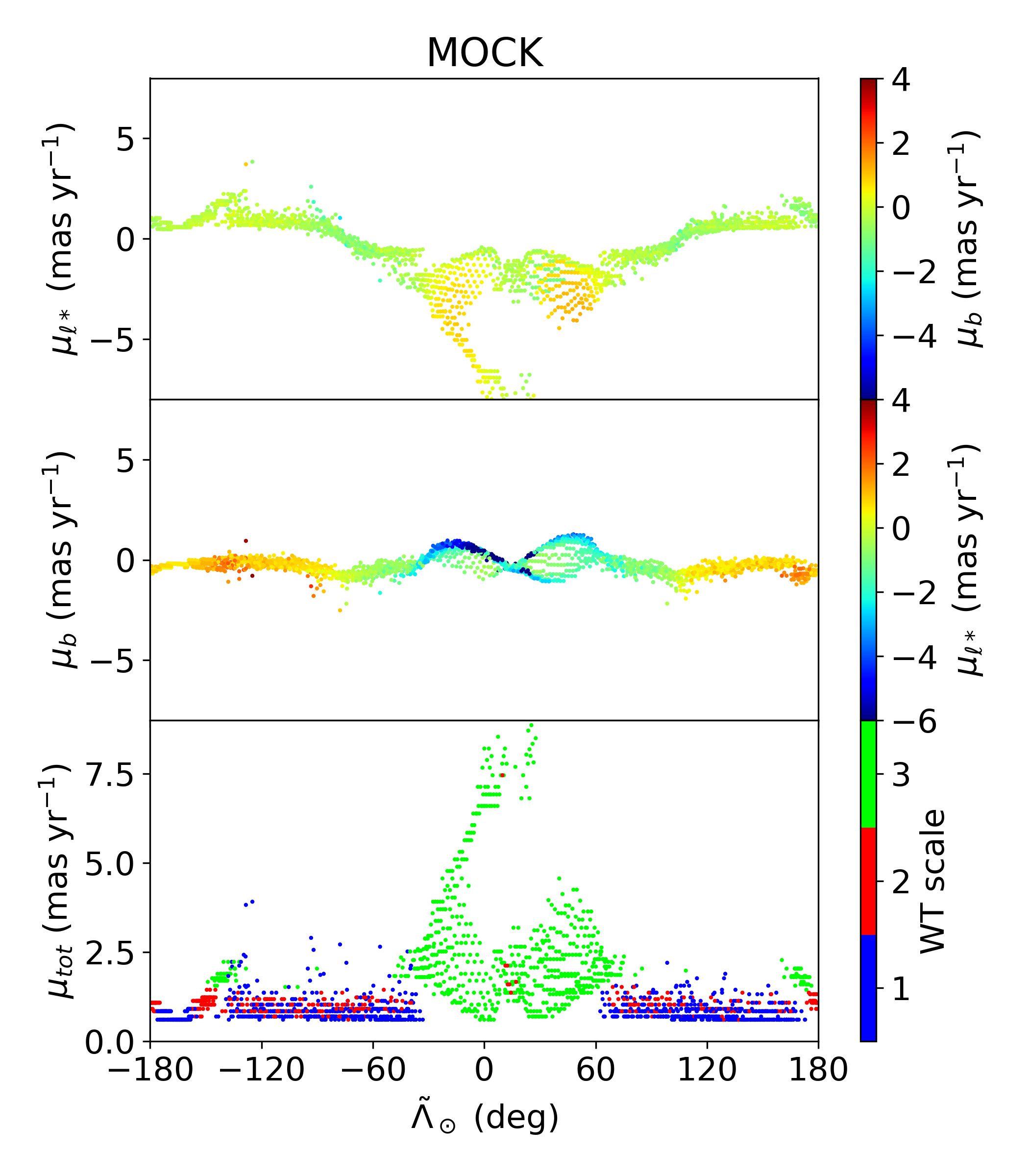}\hspace{-0.1cm}
   \includegraphics[width=0.24\textwidth]{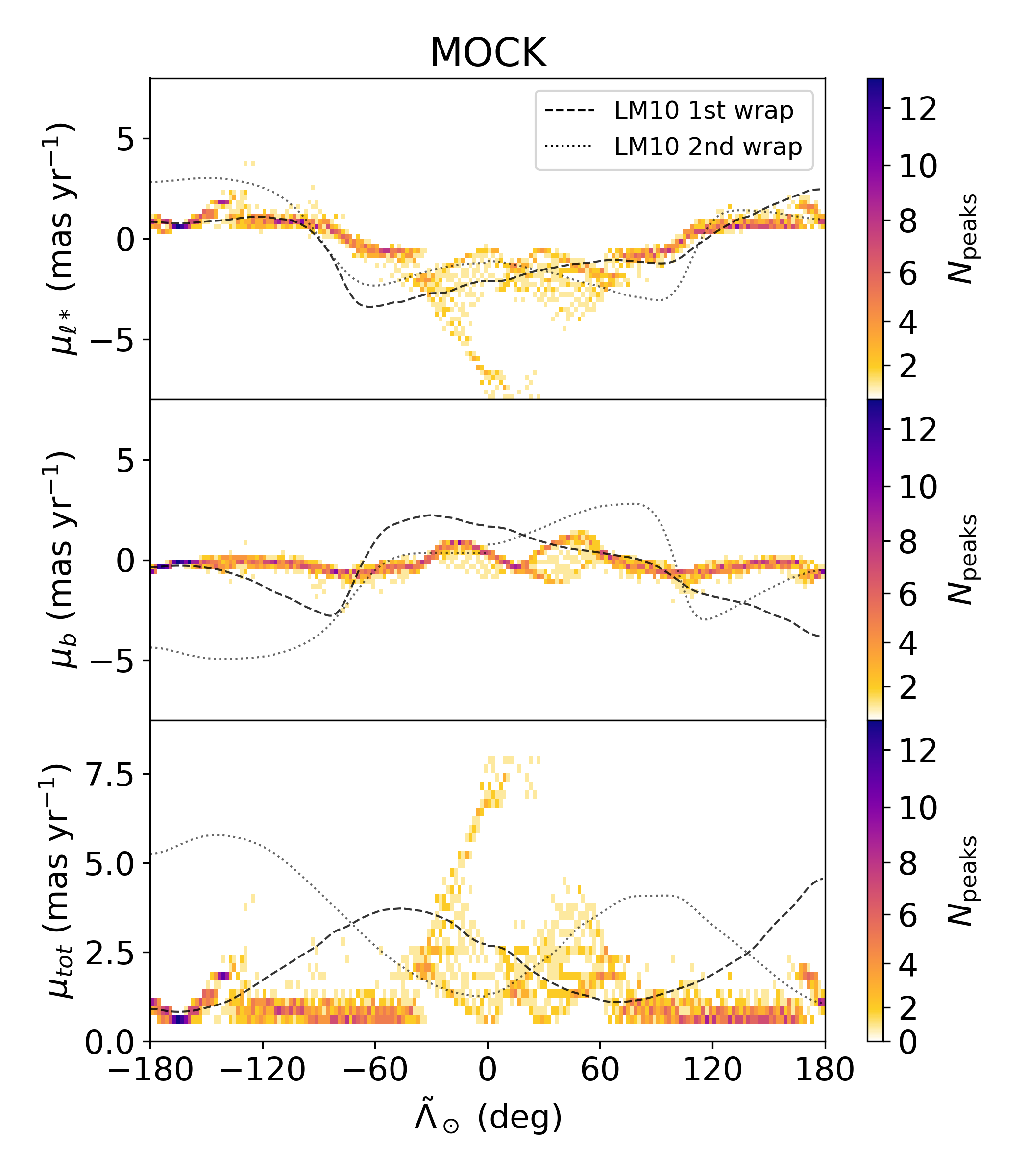}

      \caption{Peaks detected in the mock \Gaia catalogue with no Sagittarius stream. Left: Proper motions of the peaks as a function of $\L$ (top: proper motion in Galactic longitude, middle: in latitude, bottom: total proper motion) colour coded by proper motion in latitude (top), in longitude (middle), and by WT scale (bottom). Right: Same as in the left but in a two-dimensional histogram with the LM10 model superposed with black lines.}
         \label{mock} 
   \end{figure}
   
\begin{figure}
   \centering
         \includegraphics[width=0.24\textwidth]{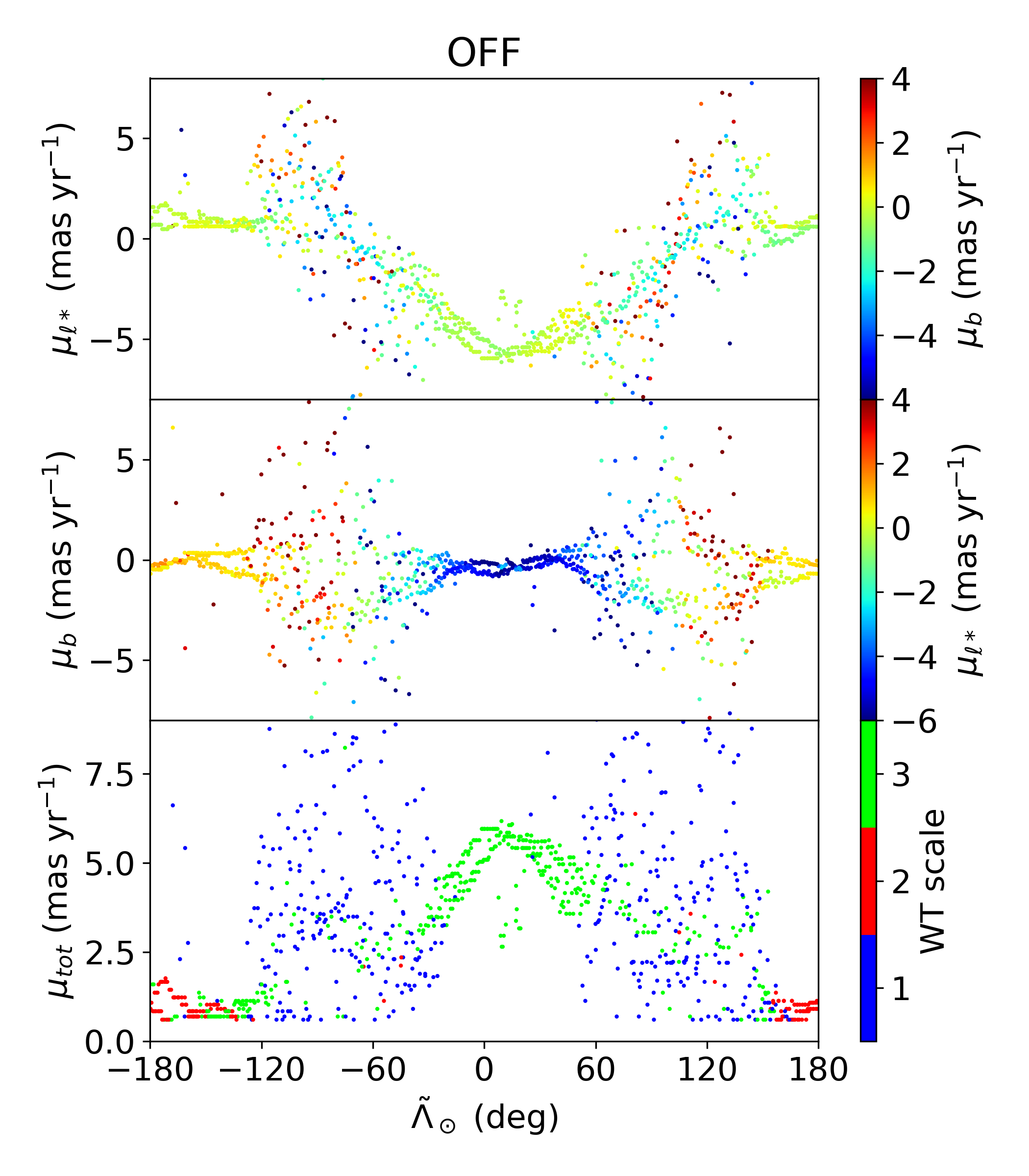}\hspace{-0.1cm}
   \includegraphics[width=0.24\textwidth]{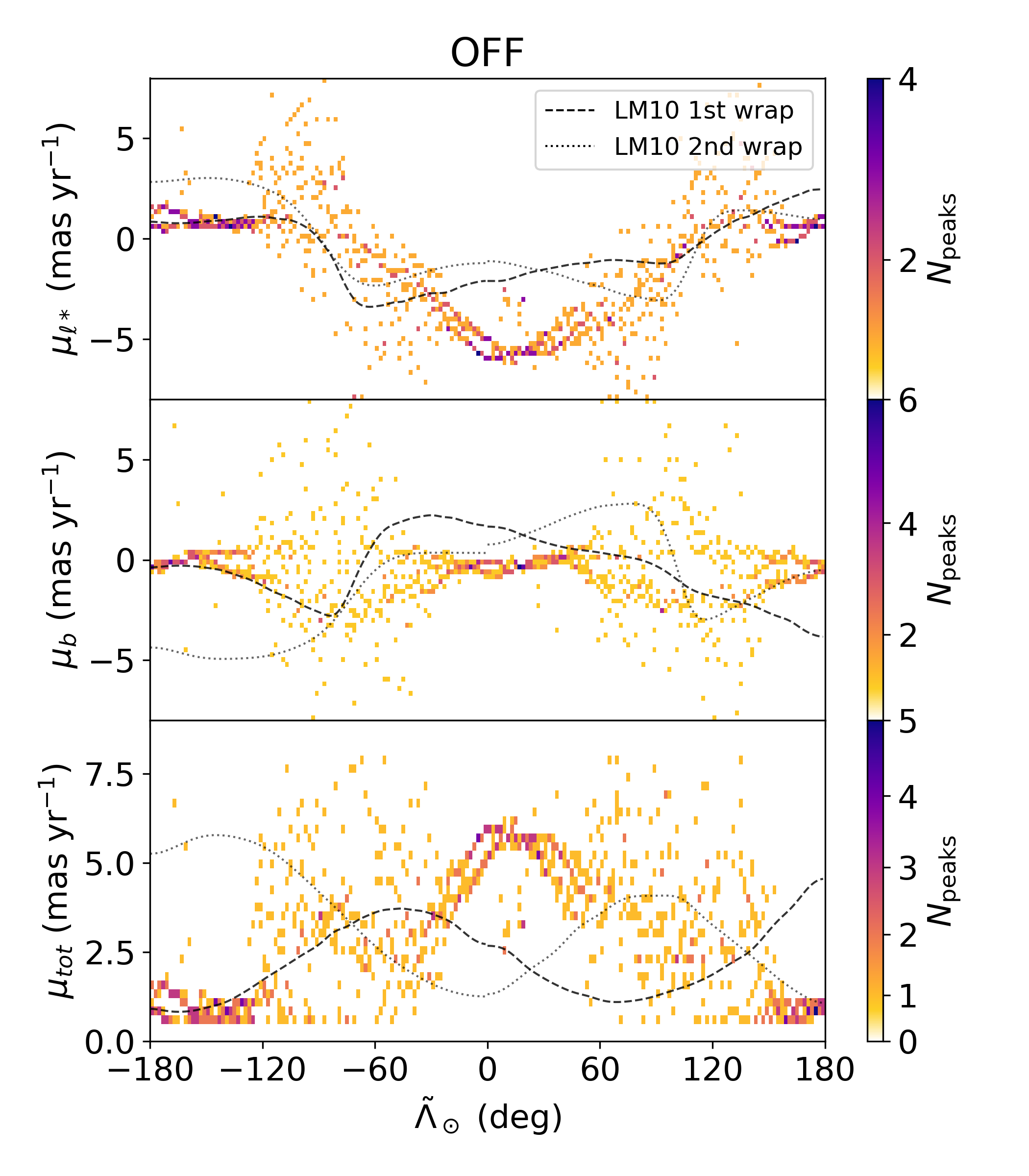}

      \caption{Same as Fig.~\ref{mock} but for the fields off-stream in the \Gaia DR2 data.}
         \label{off} 
   \end{figure}   

Figure \ref{mock} shows the peaks detected in a mock sample that does not contain the Sgr stream. 
To built this simulated catalogue we took the mock data from \citet{Rybizki2018} and added Gaussian errors to colour, magnitude, parallax and proper motions ($\mua$, $\mud$) with the uncertainties quoted. We select stars in the same manner as for the real DR2 data and we apply the same method to detect peaks in proper motion.

The mock data presents a sequence of peaks evolving with $\L$ which most likely correspond to the centre of the main component/s of the distribution (mostly foreground stars) that changes with sky position due to differences in the true mean motion of the stars and to the projection in equatorial proper motions of the solar motion. In general, we see that the sequence in the mock catalogue and in the data do not match, confirming that the peaks that we assign to Sgr do not belong to the distribution of foreground stars. The exception is the group of peaks organized in a triangular shape at middle $\L$, which as explained above, have been removed for our analysis.

We also note that around $\L\approx-180\deg$ there are mock peaks that are very close to the LM10 model in all three components ($\mul$, $\mub$, $\mutot$). This is unfortunate since in this position the stream crosses the Galactic plane and, thus, we expect the contamination to be very important. This means that the peak corresponding to the stream will be hardly distinguishable from the one for the foreground population. At $\L\approx-180\deg$ the mock data also show coincidence with the model for the first wrap but only for $\mul$, which means that the stream could be distinguished unambiguously from the other component $\mub$. However, the second wrap of the trailing arm does have a proper motion similar to that of the field.

We have also run our method on a sample of stars off-stream defined as stars with $15<|\B|\leq20\deg$. The corresponding plot of peaks is shown in Fig.~\ref{off}. We note again the coincidence of peaks that are not Sagittarius but have the same proper motion at $\L\pm180\deg$, further evidencing the contamination when the stream crosses the Galactic plane. We note again the central triangular distribution of peaks that clearly do not belong to Sagittarius.


\section{Selection of the Sagittarius peaks}\label{s:selection}

The selection of peaks of the different parts of the stream is done following the following empirical prescriptions. We first define the following lines, where proper motions are given in $\masyr$:

\begin{equation}
\begin{aligned}
{\mutot}_1&=2.3-1.4\sin(1.5\L({\rm rad})-0.5) \\
{\mutot}_2&=2.5-1.4\sin(1.5\L({\rm rad})-0.3) \\ 
{\mul}_3&=2.4 \\
{\mul}_4&=-4.+0.045\L \\
{\mub}_5&=1.2-(1./4.\times0.006)(\L+14.)^2 \\
{\mul}_6&=2.4-0.03\L 
\end{aligned}
\end{equation}

We then select only peaks satisfying the following criteria for the first part of the trailing arm ($\L<0\deg$):
\begin{equation}
\begin{gathered}
|\mutot-{\mutot}_1|<1.1 \\
\mul<{\mul}_3  \\
\mub<{\mub}_5,
\end{gathered}
\end{equation}
\noindent for the trailing small part at $\L>150\deg$: 
\begin{equation}
\begin{gathered}
|\mutot-{\mutot}_2|<-1.,
\end{gathered}
\end{equation}
\noindent and for the leading part ($\L>0\deg$ ):
\begin{equation}
\begin{gathered}
|\mutot-{\mutot}_2|<0.7  \\
\mul<{\mul}_4 {\rm \,for\,} \L<85\deg \\
\mub<{\mub}_6 {\rm \,for\,} \L>115\deg
\end{gathered}
\end{equation}

 After applying the above geometric filters and examining the proper motion in equatorial coordinates of the selected peaks,  there is a peak that is clearly off the Sgr track in $\mua$ (while our selection is done in $\mul$-$\mub$). This is caused by an abrupt change of proper motion on certain parts of the celestial sphere related to the coordinate transformation. Thus, we further remove this peak (\HP 4491, ICRS level 5).

\section{Removing stars from globular clusters}\label{s:globular}


From the peaks high relative intensities, the one for the globular clusters Pal\,12, Terzan\,7, Terzan\,8, Arp\,2, M\,54, NGC\,2419, M\,53 (NGC\,5024) and NGC\,5053 coincide with the projected path of the Sgr stream and, more importantly, with the proper motion sequence shown in the Fig. 3.  We remove the mentioned globular clusters in Fig.~\ref{cmd} and from the list of candidate members by filtering out stars in these clusters. For this we consider the positions, tidal radius and distance from \citet{Baumgardt2018}, and take the apparent size of the clusters as 5 times the tidal radius, except for NGC5024 and NGC6715 for which we use factors of 10 and 25, respectively.

\section{Candidate members table}

 Table~\ref{table:1} shows the first 5 rows of the Table in CDS {\bf (link)} containing the data of the candidate members of the Sagittarius stream delivered in this study. Please, note that these members have been selected based on proper motion only and that a certain fraction of contamination is expected.
 
\begin{table*}
\caption{First 5 rows of the table provided in CDS with the list of candidate members of the Sagittarius stream and dwarf. {\bf to be filled in}}             
\label{table:1}      
\centering          
\end{table*}

\end{appendix}

\end{document}